\documentclass[prb,aps,twocolumn,showpacs]{revtex4-1}

\usepackage{graphicx}
\usepackage{latexsym}
\usepackage{amsmath}
\usepackage{bm}

\newcommand{\be}{\begin{eqnarray}}
\newcommand{\ee}{\end{eqnarray}}

\begin{document}

\title{Phase Structure of Repulsive Hard-Core Bosons in a Stacked 
Triangular Lattice
}

\date{\today}

\author{Hidetoshi Ozawa} 
\author{Ikuo Ichinose}
\affiliation{%
Department of Applied Physics, Nagoya Institute of Technology,
Nagoya, 466-8555 Japan}

\begin{abstract}
In this paper, we study phase structure of a system of
hard-core bosons with a nearest-neighbor (NN) repulsive interaction
in a stacked triangular lattice.
Hamiltonian of the system contains two parameters one of which is
the hopping amplitude $t$ between NN sites and the other is the NN
repulsion $V$.
We investigate the system by means of the Monte-Carlo simulations
and clarify the low and high-temperature phase diagrams.
There exist solid states with density of boson $\rho={1 \over 3}$ and
${2\over 3}$, superfluid, supersolid and phase-separated state.
The result is compared with the phase diagram of the two-dimensional
system in a triangular lattice at vanishing temperature.
\end{abstract}

\pacs{67.85.Hj, 75.10.-b, 03.75.Nt}

\maketitle

In recent years, 
cold-atomic systems are one of the most intensively studied topics not only in
atomic physics but also in condensed matter physics.
In particular, study on cold-atom systems put on an optical lattice may 
give very important insight about dynamics of strongly-correlated 
particle systems\cite{optical}.
For systems in an optical lattice, interactions between atoms, 
dimensionality of system, etc. are highly controllable,
and effects of impurities are strongly suppressed.
Among many interesting topics, possibility of the appearance of the
supersolid (SS) has been investigated intensively.
In the SS, the superfluidity and a diagonal long-range order (DLRO) 
such as density wave coexist.
It has been clarified that the SS cannot exist in a hard-core (HC) boson system
in a square lattice unless long-range interactions between bosons are
included\cite{square}.
On the other hand on a triangular lattice, it was shown that the SS can
appear as a result of the competition between the particle hopping and 
nearest-neighbor (NN) repulsion\cite{triangle,squarefru}. 

In this paper, we shall pursue the above problem, i.e., 
we shall consider a three-dimensional (3D) hard-core (HC) boson system in a stacked 
triangular lattice and study a finite-temperature ($T$) phase diagram.
As we consider the 3D system, there exist finite-temperature ($T$) phase
transitions in addition to ``quantum phase transition".
Hamiltonian of the system is then given by
\begin{eqnarray}
H &=&-t\sum_{\langle i,j\rangle}(\phi^\dagger_i\phi_j+\phi^\dagger_j\phi_i) 
\nonumber \\
 && +V\sum_{\langle\langle i,j\rangle\rangle}n_in_j-{\mu}\sum_i n_i,
\label{H1}
\end{eqnarray}
where $i$ and $j$ denote site of 3D stacked triangular lattice, 
$\phi_i$ is annihilation operator of the HC boson,
$n_i$ is the number operator $n_i=\phi^\dagger_i \phi_i$ and $\mu$ is the chemical potential for the 
grand-canonical ensemble.
$\langle i,j \rangle$ denotes the NN sites in the 3D stacked triangular lattice, whereas
$\langle\langle i,j\rangle\rangle$ those of the 2D triangular lattice.
$t$ is the hopping amplitude in the 3D lattice, whereas $V(>0)$
denotes the NN repulsion in the 2D triangular lattice.
The HC boson operators $\phi_i$ satisfy the following (anti)commutation relations,
\begin{eqnarray}
&& [\phi_i,\phi_j]=[\phi^\dagger_i,\phi^\dagger_j]=[\phi^\dagger_i, \phi_j]=0,
\;\; \mbox{for} \;\; i\neq j, \nonumber \\
&& \{\phi_i, \phi^\dagger_i\}=1, \; \{\phi_i, \phi_i\}=0, \;
\{\phi^\dagger_i, \phi^\dagger_i\}=0.
\end{eqnarray}
The above HC boson can be expressed in terms of 
the Schwinger boson $w_{i\sigma}$ ($\sigma=1,2$),
\begin{equation}
\phi^\dagger_i=w^\dagger_{i1}w_{i2},
\label{Sboson}
\end{equation}
where the physical-state condition of the Schwinger is given as,
\begin{equation}
\Big(\sum_{\sigma=1,2}w^\dagger_{i,\sigma}w_{i,\sigma}-1\Big)
|{\rm Phys}\rangle_w =0,
\end{equation}
where $|{\rm Phys}\rangle_w$ is the physical Hilbert space of $w_{i\sigma}$.

Partition function of the system at temperature ($T$)
$Z={\rm Tr} \exp(-\beta H)$ with $\beta=1/k_{\rm B}T$,
is given as follows,
\begin{eqnarray}
Z&=&\int \prod_{i,\tau}[dw_i(\tau)]\exp \Big( \int^\beta_0 d\tau A(\tau)\Big),
\nonumber \\
A(\tau)&=&-\sum_i \bar{w}_i\dot{w}_i(\tau)-H(w(\tau)),  
\label{Z1}
\end{eqnarray} 
where $[dw_i(\tau)]=\delta(\sum_\sigma \bar{w}_{i\sigma}w_{i\sigma}-1)
d^2w_i(\tau)$.
In the present study, we shall ignore the $\tau$-dependence of $w_{i\sigma}$,
and then focus on the finite-$T$ physical properties,
\begin{equation}
Z'=\int \prod_{i}[dw_i]\exp (-\beta H(w)).
\label{Z2}
\end{equation}
In the previous papers\cite{BtJ1,hopping}, 
we discussed that state with a LRO found in the
above approximation at finite $T$ survives to $T=0$ unless it is
replaced by another ordered state.
This expectation will be again confirmed by the study in this paper 
through the comparison between the obtained results and those of system in 
a 2D triangular lattice at $T=0$\cite{triangle}. 
From Eq.(\ref{Z2}), we introduce {\em dimensionless parameters} $c_1=\beta t$
and $c_2=\beta V$.
To obtain phase diagram, we fix the parameter $c_1$ and investigate
the system by varying values of $c_2$ and $\mu$
(the grand-canonical ensemble), or the density of $\phi_i$ (the canonical
ensemble).
These two calculations, i.e., the grand-canonical and canonical ensembles, 
are complementary with each other.
For the simulations, we employ the standard Monte-Carlo Metropolis algorithm
with local update calculation.
The typical sweeps for measurement is $(30000 \sim 50000)\times (10$ samples).
\begin{figure}[h]
\begin{center}
\includegraphics[width=7cm]{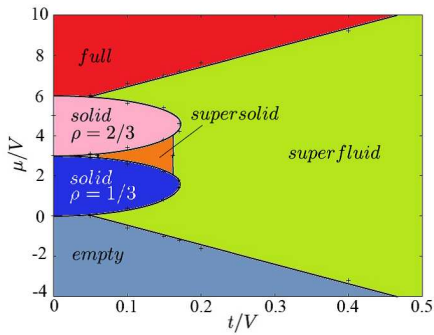}
\caption{
Phase diagram in the grand-canonical ensemble for $c_2=V/k_{\rm B}T=50.0$,
which corresponds to the low-temperature phase.
}\vspace{-0.5cm}
\label{PD1}
\end{center}
\end{figure}
\begin{figure}[t]
\begin{center}
\includegraphics[width=4.1cm]{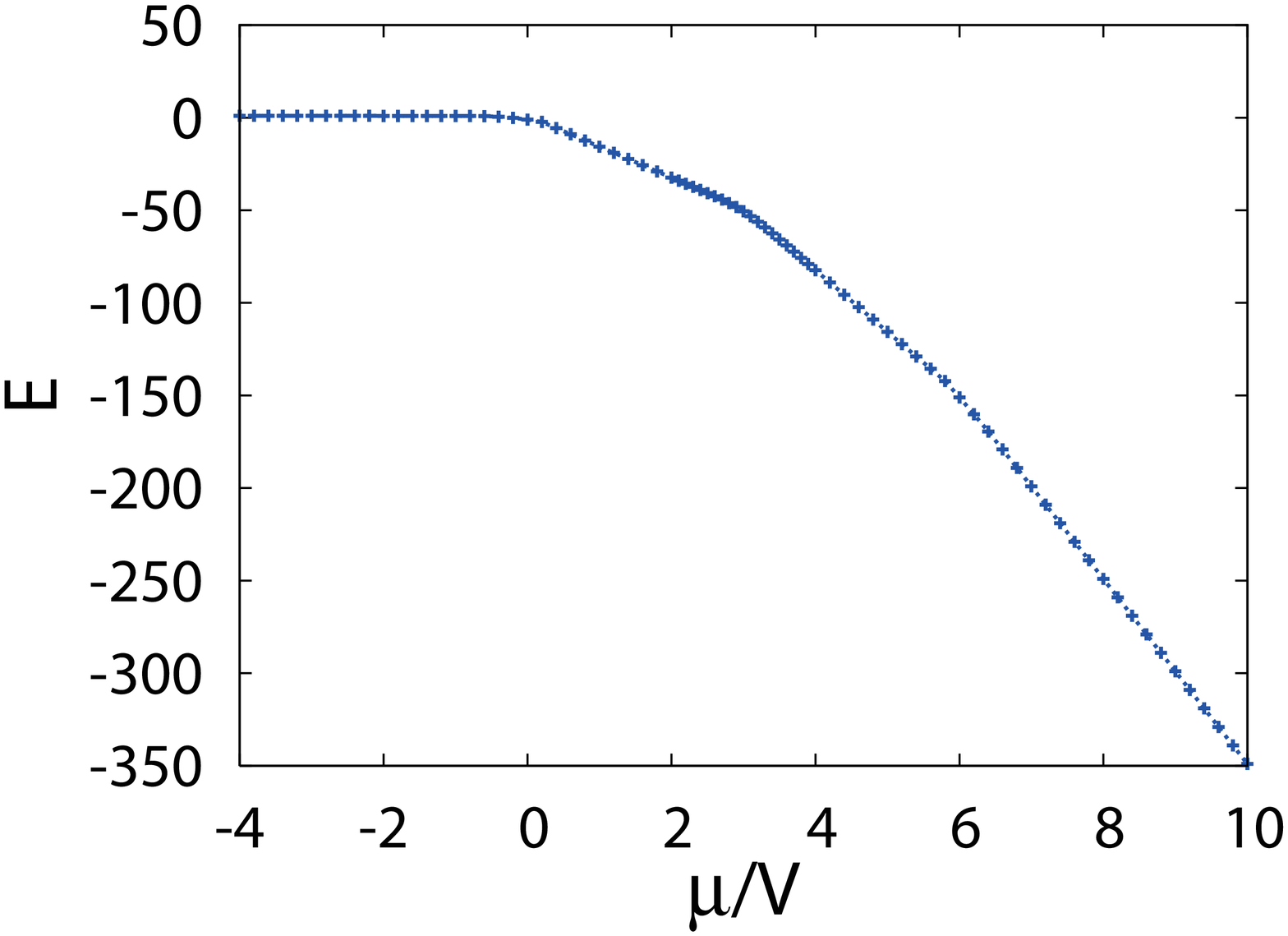}
\includegraphics[width=4.1cm]{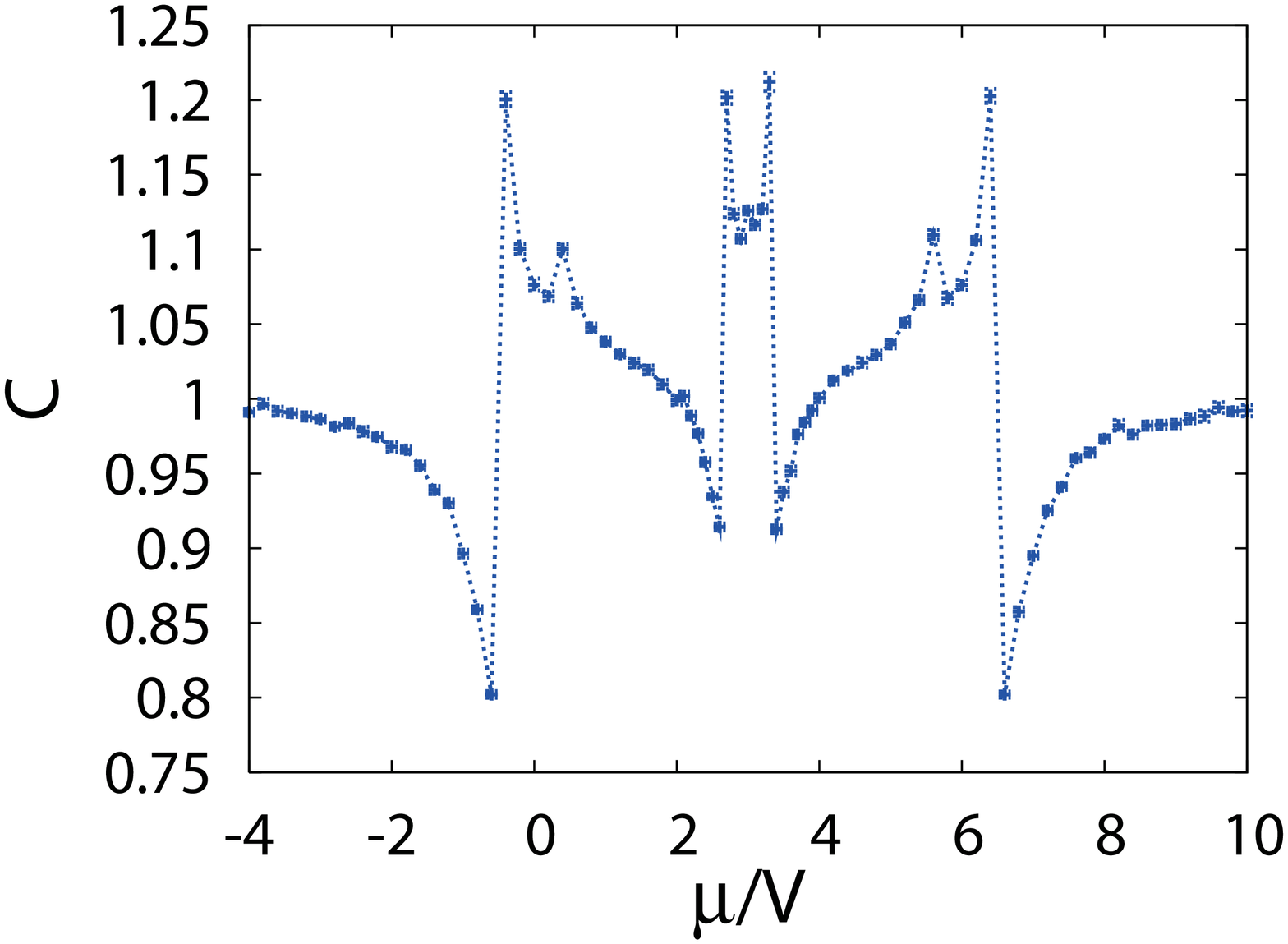}  \vspace{0.5cm} \\
\includegraphics[width=4.5cm]{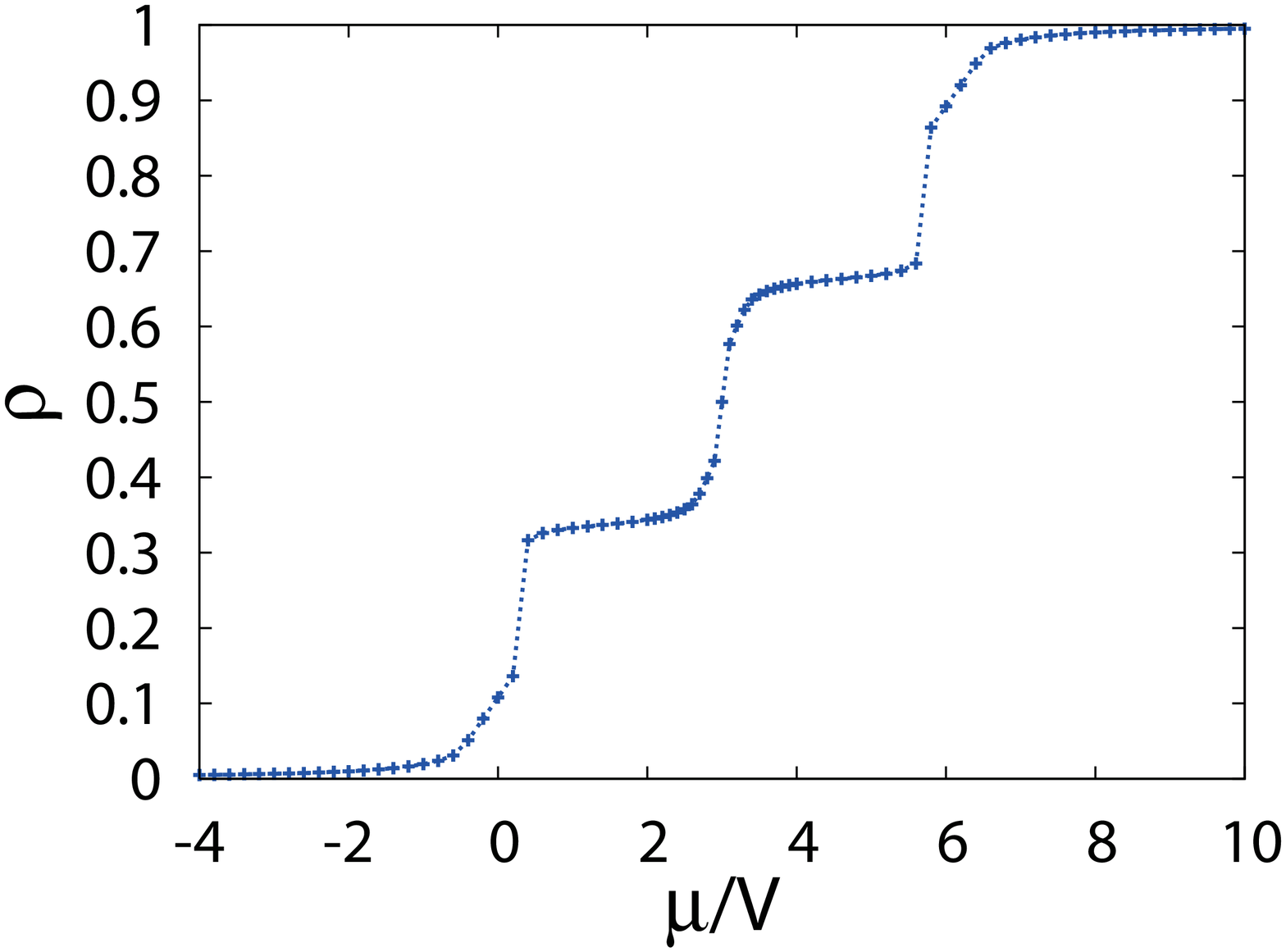}
\caption{
Internal energy, specific heat and particle density as a function of
$\mu/V$ for $c_1=5$ and $c_2=50$ $(t/V=0.1)$ in the grand-canonical ensemble.
System size $L=24$.
}
\label{grand1}
\end{center}
\end{figure}

We first show the phase structure of the system in the grand-canonical
ensemble.
To investigate the phase structure, we measured the internal energy $E$,
the specific heat $C$, the particle density $\rho$, etc, which are
defined as $E=\langle H \rangle/L^3$, $C=\langle (H-E)^2 \rangle/L^3$
and $\rho=\langle n_i\rangle$, where $L$ is the linear system
size.
We numerically studied the system at various $T$'s and found that
the phase diagram changes continuously.
For the low-$T$ phase, we set $c_2=50$ and obtain the phase diagram
in the $t/V-\mu/V$ plane.
See Fig.\ref{PD1}.
There are full and empty states in the phase diagram, in which all sites of the system 
are occupied or empty, respectively.
For $t/V \ll 1$, there appear ``solid" states with the density $\rho={1 \over 3}$
and $\rho={2 \over 3}$.
It is easily seen that these states in which one of three sites is filled (empty)
in a $\sqrt{3}\times \sqrt{3}$ ordering have minimum energy for 
$t/V \ll 1$\cite{Met}.
As  $t/V$ is increased, there appear the SS and superfluid (SF) states.
The obtained phase diagram in Fig.\ref{PD1} is essentially the same with
that of the system in the 2D triangular lattice at $T=0$\cite{triangle}.
The three dimensionality of the present system preserves the ordered states
at $T=0$ in 2D up to the low but finite $T$.

In the following, we shall exhibit the MC calculations that
derived the above phase diagram in Fig.\ref{PD1}.
First in Fig.\ref{grand1}, we show $E$, $C$ and $\rho$ as a function
of $\mu$ for $c_1=5$ and $c_2=50$, i.e., $t/V=0.1$.
It is obvious that there is a particle-hole symmetry.
Singular behavior of $C$ indicates that there are totally six phase transitions
and the step-function like behavior of
$\rho$ indicate that phase transitions at $\mu/V\simeq 0, 3$
and $6$ are of first order.

\begin{figure}[h]
\begin{center}
\includegraphics[width=4cm]{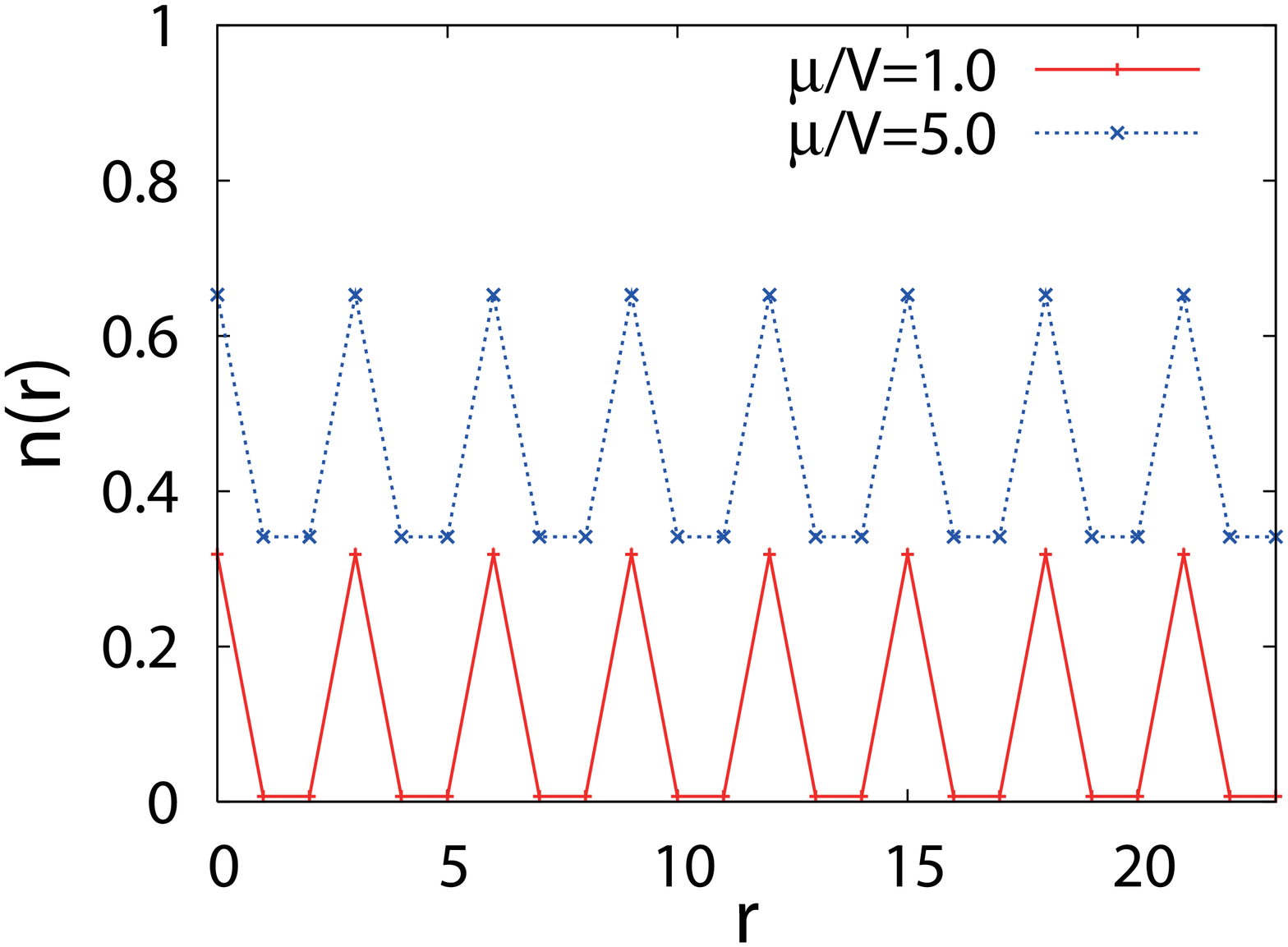}
\includegraphics[width=4cm]{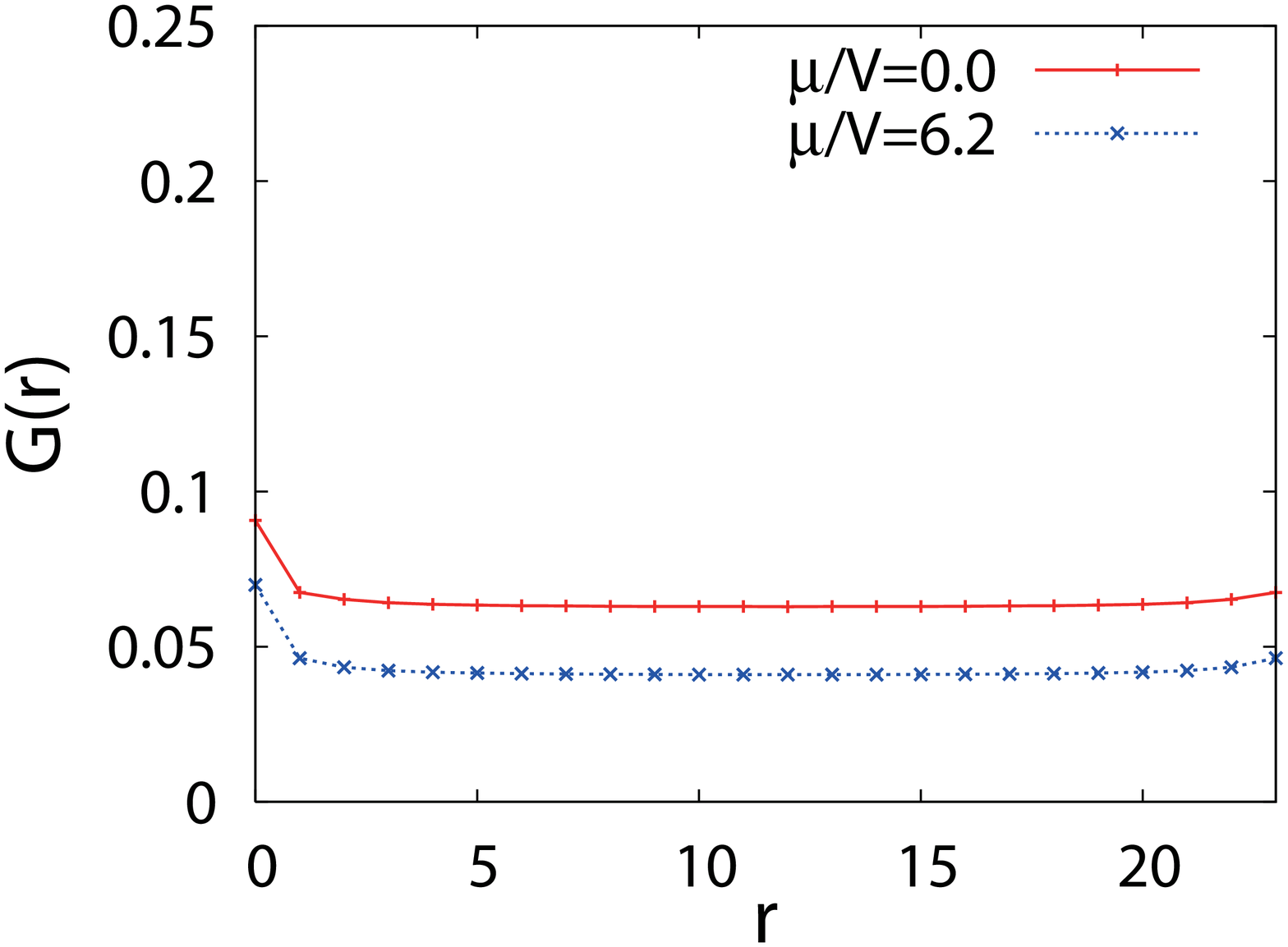} \vspace{0.5cm} \\
\includegraphics[width=4cm]{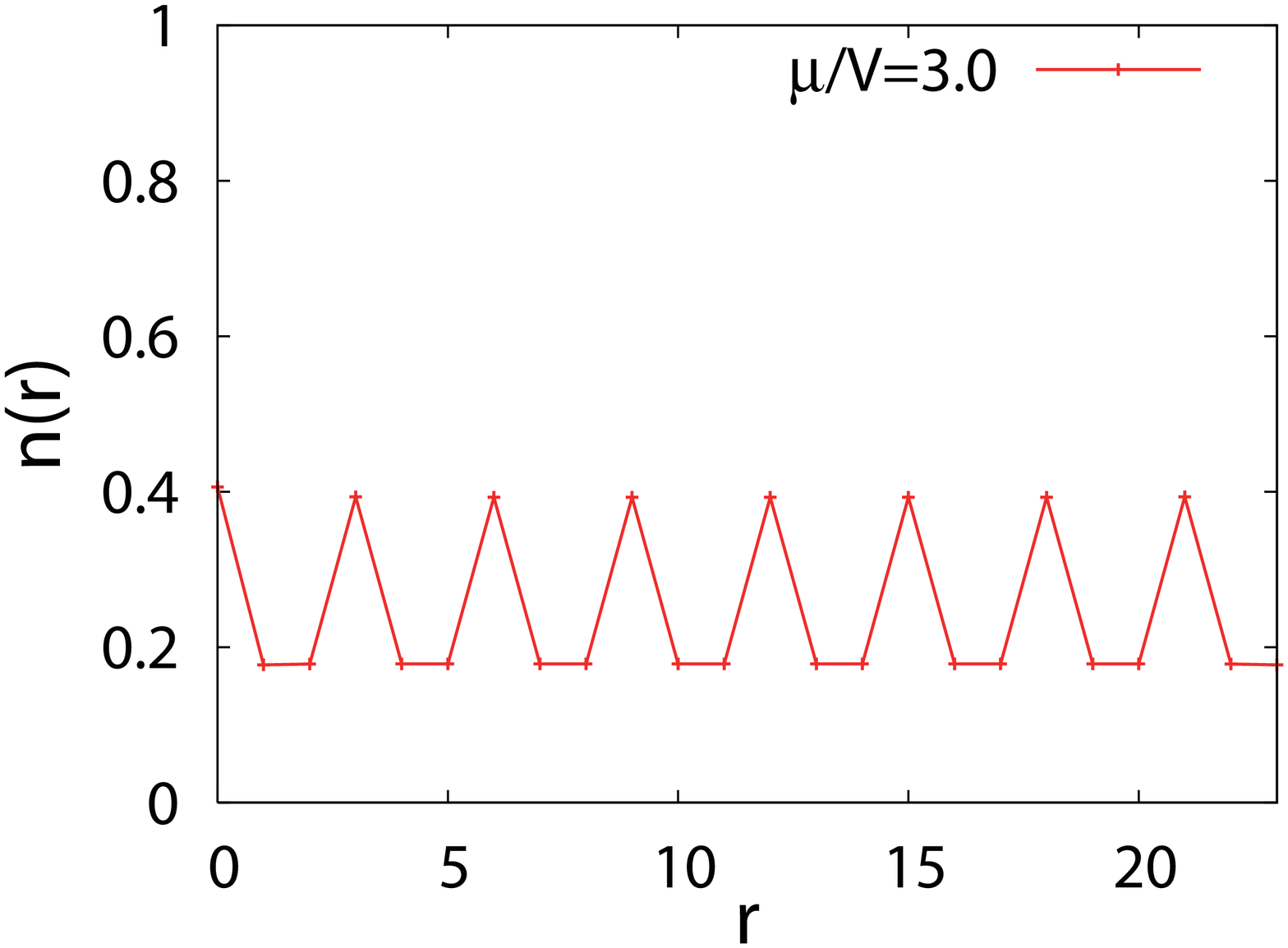}
\includegraphics[width=4cm]{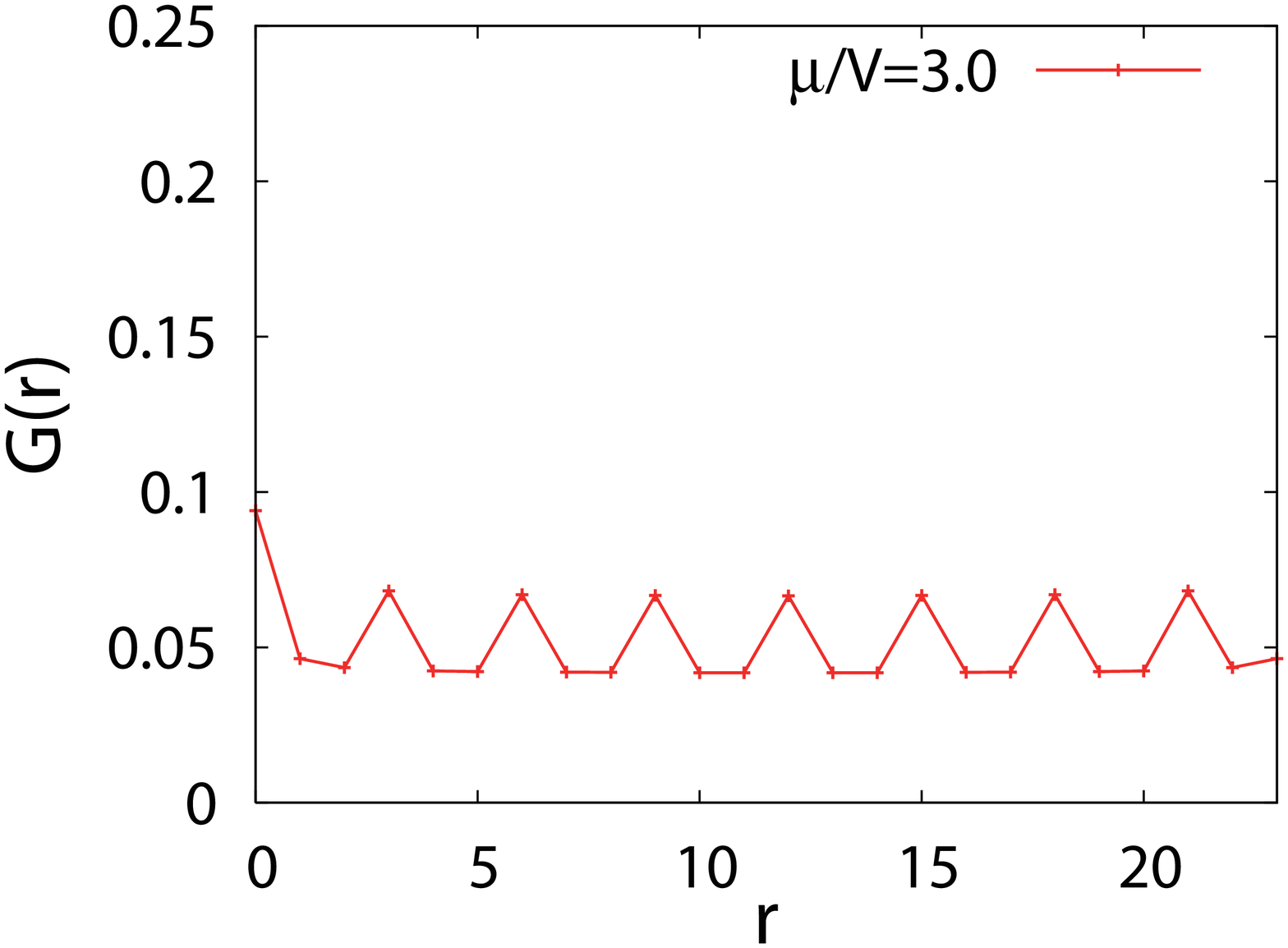}
\caption{
Correlation functions $n(r)$ and $G(r)$ in various phases
for $c_1=5$ and $c_2=50$ $(t/V=0.1)$ in the grand-canonical ensemble.
$L=24$.
}\vspace{-0.7cm}
\label{corr_grand}
\end{center}
\end{figure}

In order to verify the physical meaning of each phase,
we measured the particle-density correlation function $n(r)$ and 
the HC boson correlation function $G(r)$, which are defined as follows,
\begin{eqnarray}
n(r)=\langle \phi^\dagger_i\phi_i\phi^\dagger_{i+r}\phi_{i+r}\rangle, \;\;
G(r)=\langle \phi^\dagger_i \phi_{i+r} \rangle.
\label{corF}
\end{eqnarray}
It is expected that in the solid states with $\rho={1\over 3}$ and ${2\over 3}$,
$n(r)$ exhibits a specific correlation with $3a$ periodicity ($a=$ lattice spacing),
whereas vanishing superfluid
correlation, i.e., $G(r) \rightarrow 0$ for $r \rightarrow$ large.
On the other hand in the superfluid state, $n(r)$ exhibits a homogeneous distribution
of bosons and there exists a finite superfluid correlation.
Supersolid has finite correlation of the both $n(r)$ and $G(r)$.
We have verified all of the above expectations by the MC simulations,
and here we show some of the results in Fig.\ref{corr_grand}.
In the solid states, the HC bosons are localized in a specific pattern to avoid
repulsions and as a result a density wave appears.
Then, the superfluid correlation decays very rapidly there.
The superfluid is a homogeneous state with nonvanishing HC boson LRO. 
On the other hand in the SS, both the density wave and HC boson LRO appear.

\begin{figure}[t]
\begin{center}
\includegraphics[width=5cm]{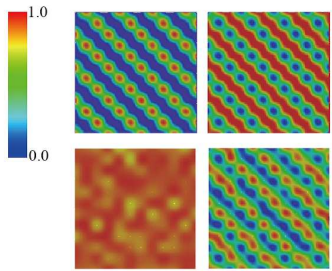}
\caption{
Snapshot of density $n(r)$ in various phases
for $c_1=5$ and $c_2=50$ $(t/V=0.1)$ in the grand-canonical ensemble.
Upper left: $\rho={1 \over 3}$ solid, Upper right: $\rho={1 \over 3}$ solid, 
Lower left: SF, and Lower right: SS.
}\vspace{-0.5cm}
\label{snap1}
\end{center}
\end{figure}

Snapshots are useful to get an intuitive physical
picture of each phase.
In Fig.\ref{snap1}, we show snapshots of particle density of each phase.
In the solid states, particles are localized, whereas in the superfluid
a homogeneous state is realized.
In the SS, the high-density region fluctuates and localized domains 
come to overlap with each other.

\begin{figure}[t]
\begin{center}
\includegraphics[width=6.5cm]{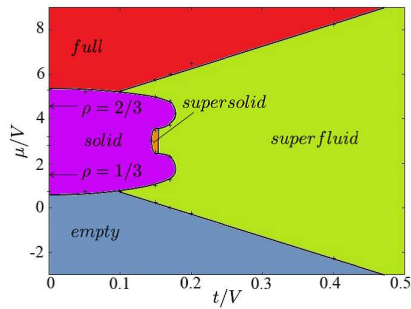}
\caption{
Phase diagram in the grand-canonical ensemble for $c_2=V/k_{\rm B}T=10.0$
}\vspace{-0.7cm}
\label{PD2}
\end{center}
\end{figure}

\begin{figure}[t]
\begin{center}
\includegraphics[width=4cm]{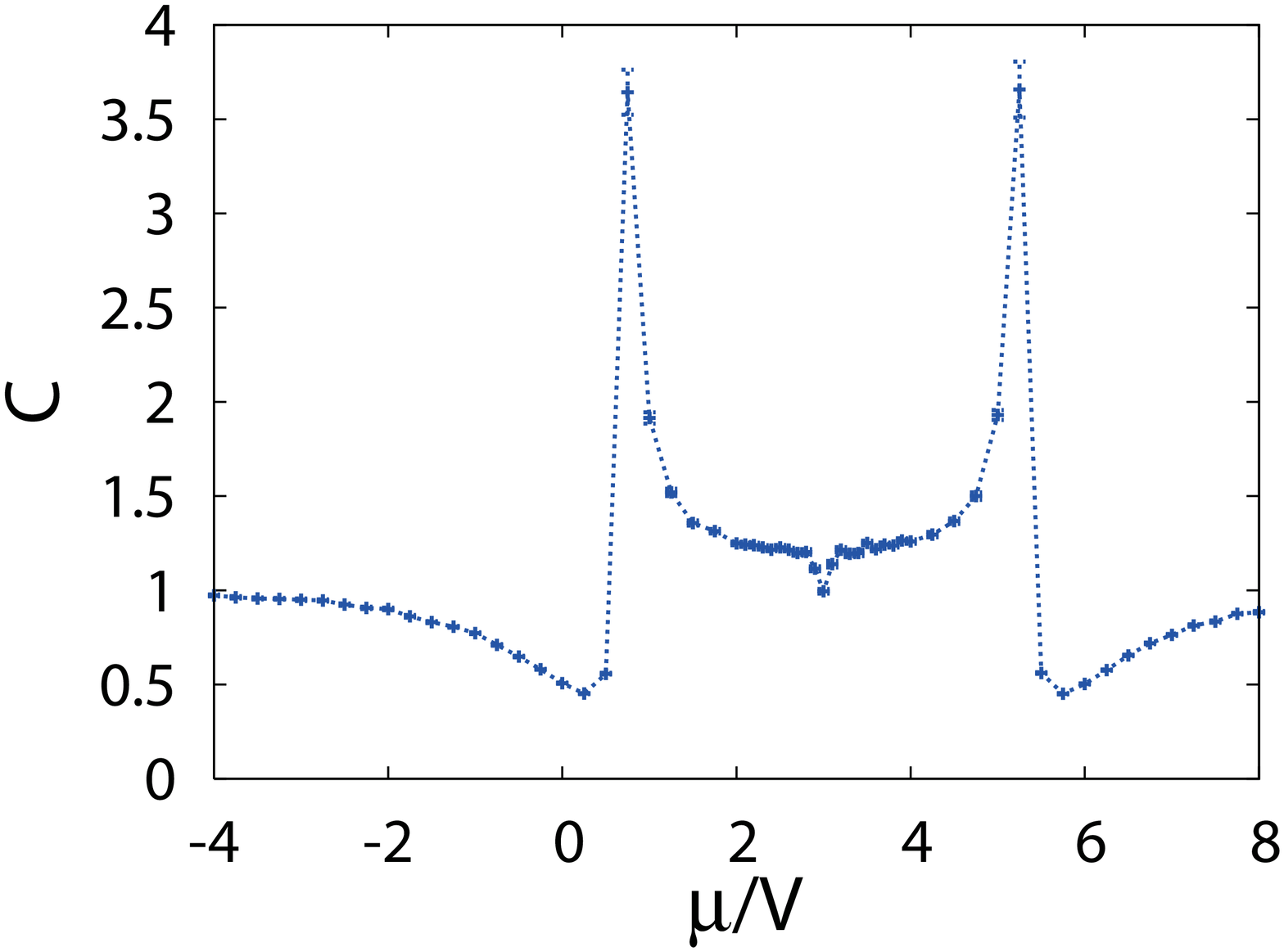}
\includegraphics[width=4cm]{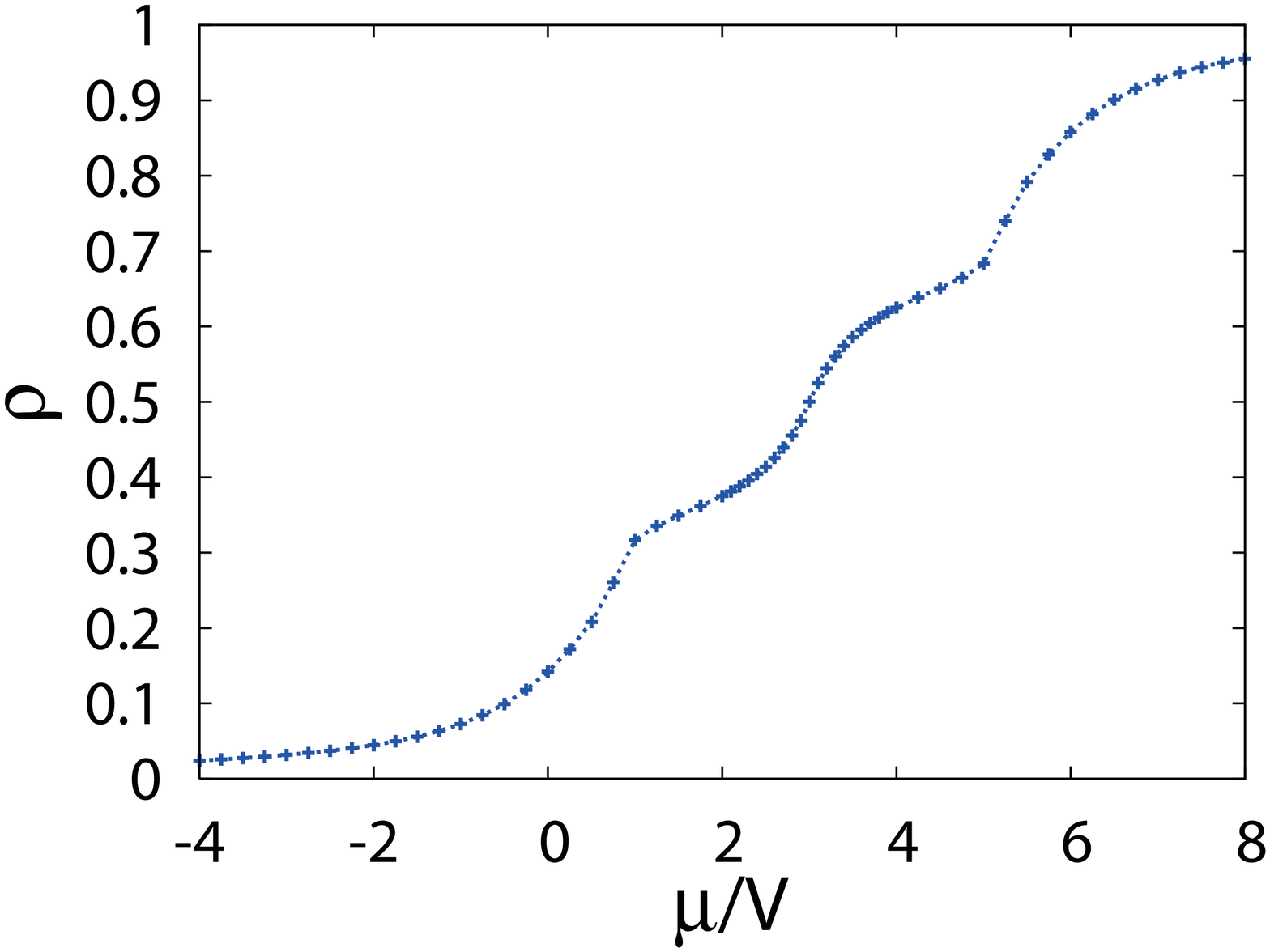} 
\caption{
Specific heat and density as a function of $\mu/V$
for $c_1=1.0$ and $c_2=10$ (i.e., $t/V=0.1$).
}
\label{Candrho_HT}
\end{center}
\end{figure}

\begin{figure}[t]
\begin{center}
\includegraphics[width=4cm]{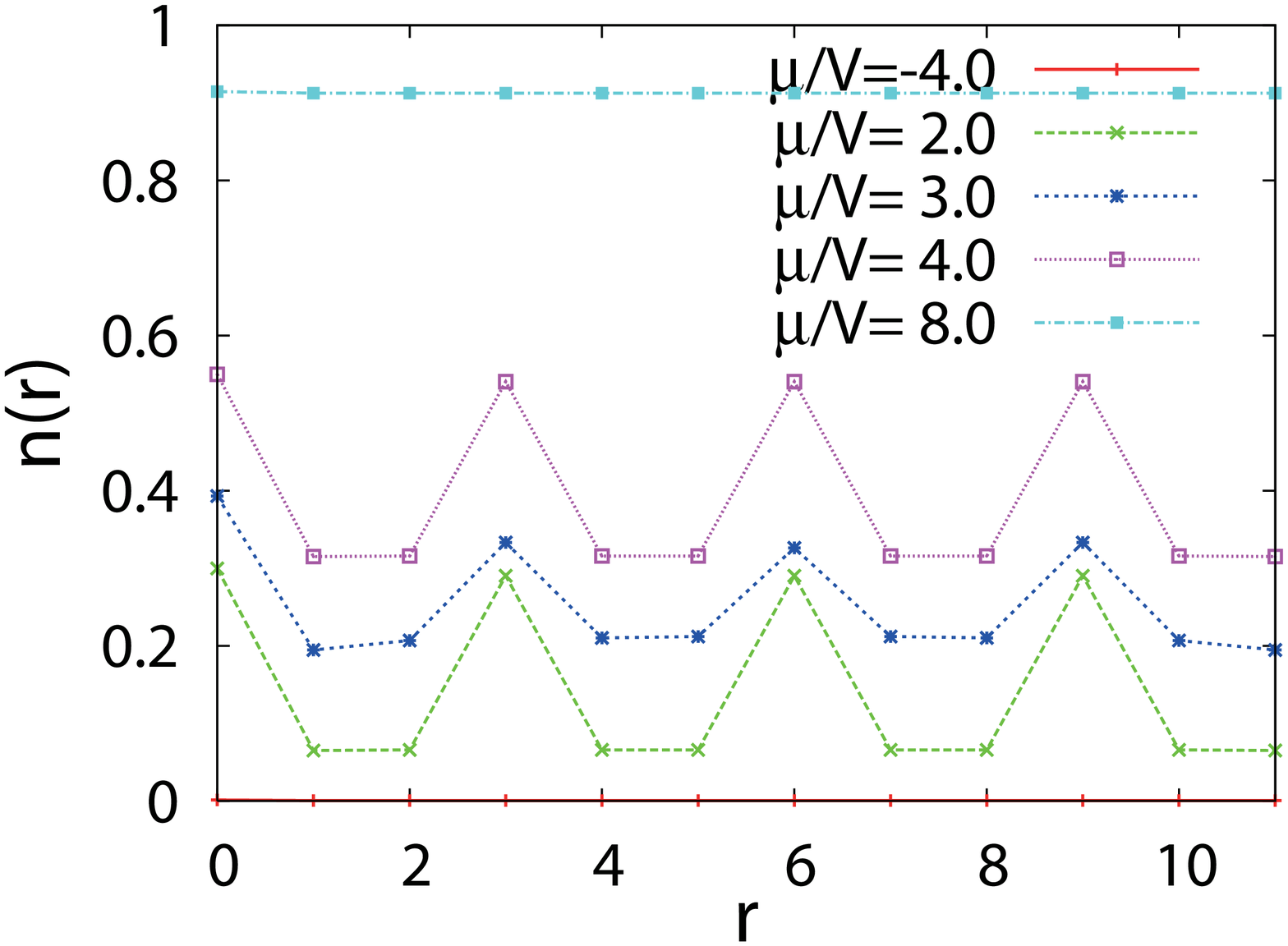}
\includegraphics[width=4cm]{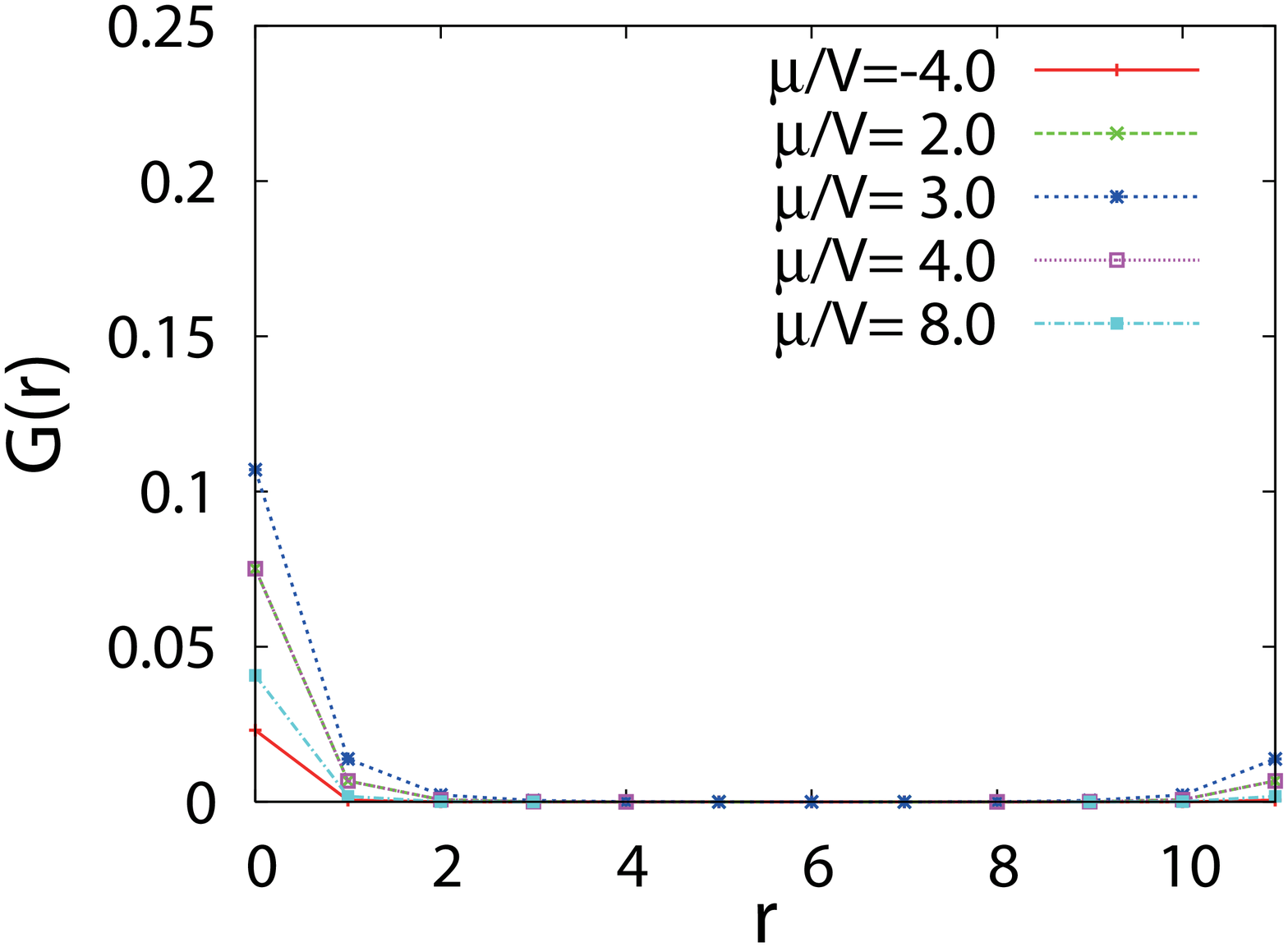} 
\caption{
Correlations of density and particle 
for $c_1=1.0$ and $c_2=10$.
}\vspace{-0.7cm}
\label{cor_HT}
\end{center}
\end{figure}

It is interesting to see how the phase diagram changes as $T$ is raised.
As we are studying the 3D system, we can address this interesting problem.
We show the phase diagram of the system with $c_2=\beta V=10.0$ in Fig.\ref{PD2}.
It is obvious that the parameter regions of the SS and also SF 
are getting small compared to those at low $T$.
Phase boundary of the two solids with $\rho={1\over 3}$ and ${2 \over 3}$
becomes obscure.
The calculated specific heat $C$ and the particle density $\rho$ are show
in Fig.\ref{Candrho_HT} as a function of the chemical potential.
For $c_1=1.0$ and $c_2=10.0$, there are two sharp peaks in $C$
at $\mu/V\simeq 1.0$ and $5.5$, which
are related with each other by the particle-hole symmetry,
and correspond to the empty-solid/solid-full phase transitions.
There also exists a small V-shape hollow at $\mu/V \simeq 3.0$, i.e., 
$\rho \simeq 0.5$.
Calculation of $\rho$ exhibits rather smooth behavior without
the sharp step-function like discontinuities that exist at low $T$.
The obtained correlation functions in Fig.\ref{cor_HT} indicate that a mixed state of
the two solids with $\rho={1 \over 3}$ and  $\rho={2 \over 3}$ appears
for $1.0<\mu/V <5.5$ and the density of each solid depends on value of  $\mu/V$.


\begin{figure}[b]
\begin{center}
\includegraphics[width=6cm]{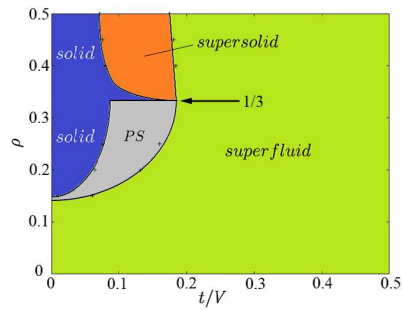}
\caption{
Phase diagram in the canonical ensemble for $c_2=50.0$.
}
\label{PD3}
\end{center}
\end{figure}
\begin{figure}[t]
\begin{center}
\includegraphics[width=5cm]{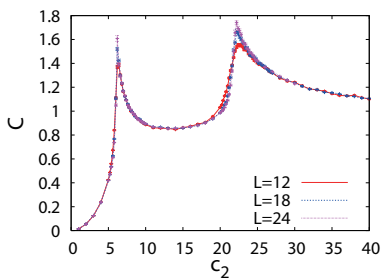}
\caption{
Specific heat $C$ for $t/V=0.1$ and $\rho=0.5$ as a function of $c_2$.
There exist two peaks indicating two second-order phase transitions.
}
\label{SS-T}
\end{center}
\end{figure}
\begin{figure}[t]
\begin{center}
\includegraphics[width=4cm]{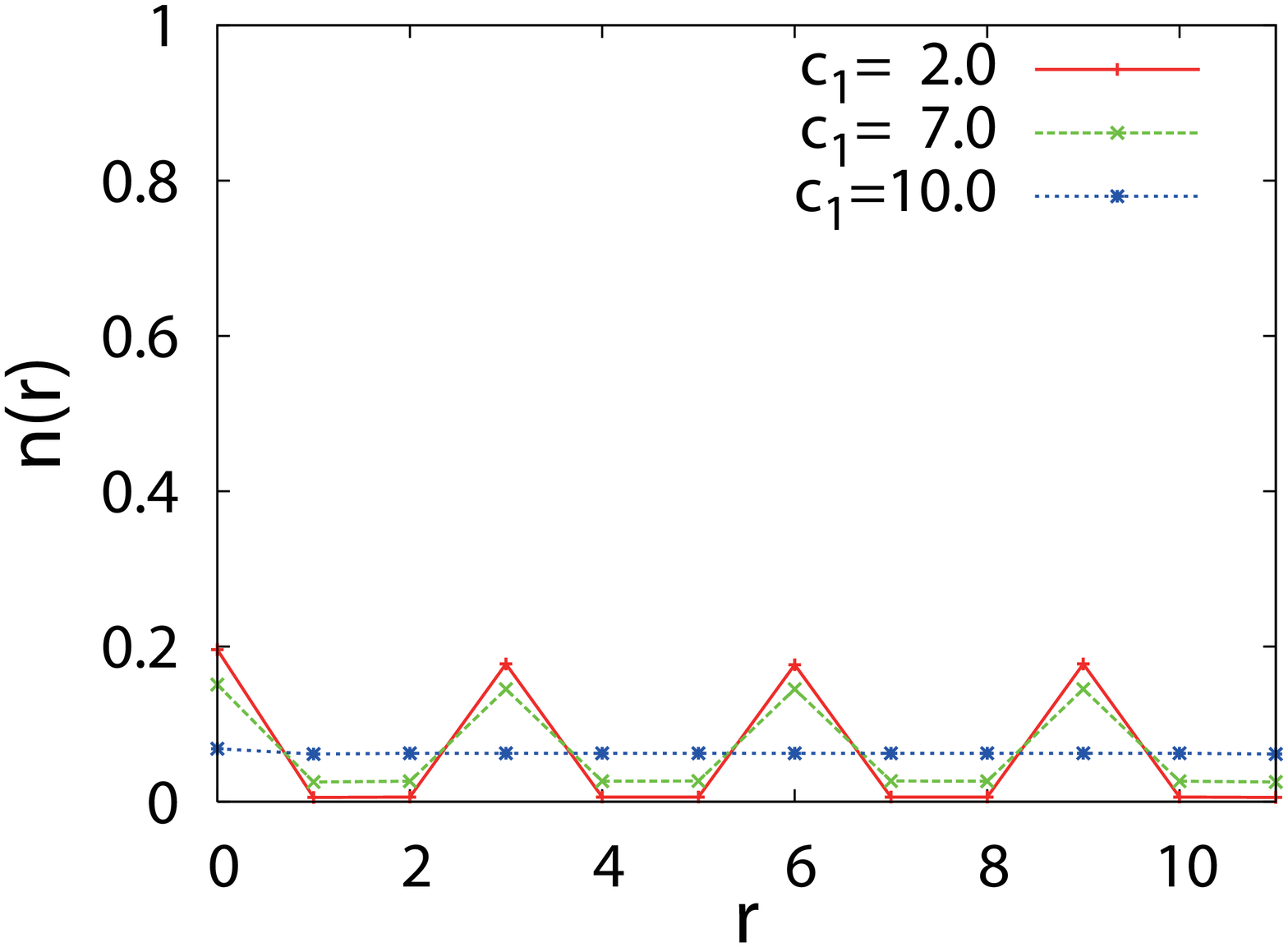}
\includegraphics[width=4cm]{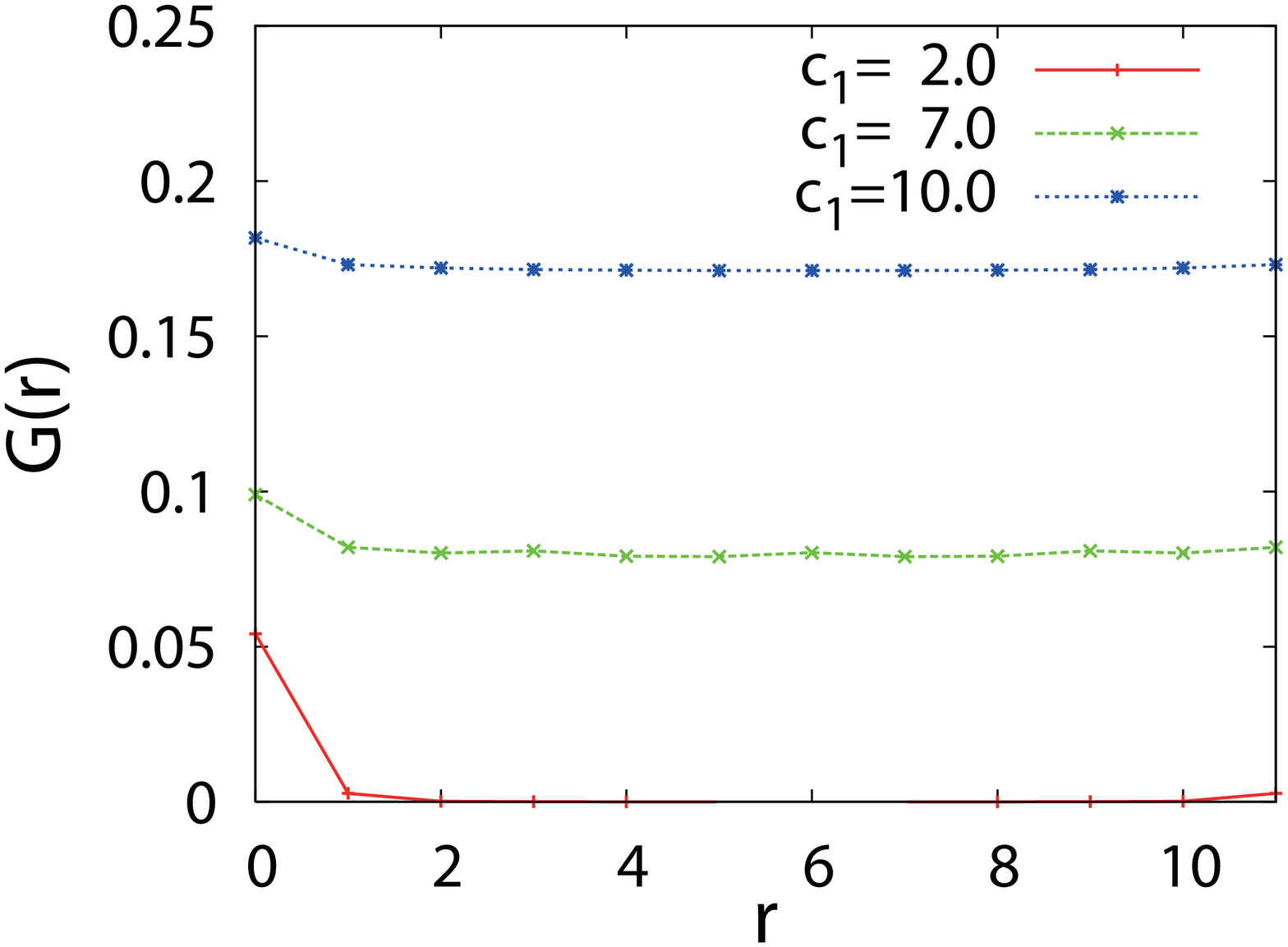}
\caption{
Number and HCB correlation functions in the canonical ensemble for $c_2=50.0$
and $\rho=0.25$.
}\vspace{-0.5cm}
\label{cor_canonical}
\end{center}
\end{figure}

Let us turn to the phase diagram of the system in the canonical ensemble.
We use the MC simulations with local updates that conserve the total
boson number.
We consider the low-$T$ case with $c_2=50$ and show the phase diagram
in the $t/V-\rho$ plane in Fig.\ref{PD3}.
We show only the region $0<\rho <0.5$ because the particle-hole symmetry 
gives the other half of the parameter region.
For small $t/V$ and $0.15<\rho <0.85$, the system exists in the solid state.
From the above study on the system in the grand-canonical ensemble,
this state can be regarded as a mixture of the solids with $\rho={1 \over 3}$
and ${2 \over 3}$.
From $\rho=0.33$ to 0.67 and $0.1 < t/V <0.2$, the SS appears as in the 
grand-canonical ensemble.
This result seems to be in sharp contrast with the phase diagram in the 2D triangular
lattice at $T=0$, in which the SS appear even for very small $t/V$\cite{triangle}.
This apparent discrepancy is closely related with 
how the SS evolves as $T$ is raised.
To see this, we study the system with $t/V=c_1/c_2=0.1$ and $\rho=0.5$
by varying the value of $\beta$.
See Fig.\ref{SS-T}, in which the specific heat $C$ is shown as a function
of $c_2$.
The result shows that
there exist two phase transitions at $c_2\simeq 8.0$ and $22$ both of which are
of second order.
By studying the correlation functions, we found that as $T$ is raised, the phase
transition from the SS to the solid state takes place first at $c_2\simeq 22$ and then
at  $c_2\simeq 8.0$ to the state without any long-range orders.
See also Figs.\ref{PD1} and \ref{PD2}.
Therefore the obtained finite-$T$ phase diagram is consistent with 
that at $T=0$ in 2D.
It should be remarked that in {\em the 2D system at finite-$T$ with
$\rho=0.5$}, corresponding phase transitions take place almost simultaneously\cite{2DT}.
In Fig.\ref{cor_canonical}, we show the correlation functions $n(r)$ and $G(r)$
for $c_2=50.0$ and $\rho=0.25$ in the canonical ensemble.

In the phase diagram in the canonical ensemble shown in Fig.\ref{PD3}, 
the phase-separated (PS) state 
also exists, which does not appear in the grand-canonical ensemble.
Similar result was obtained for the system of the 2D triangular lattice at
$T=0$\cite{triangle}.
In the grand-canonical ensemble, the most stable state appears
for each value of the chemical potential, and an inhomogeneous
state is sometimes difficult to be observed.

\begin{figure}[h]
\begin{center}
\vspace{0.3cm}
\includegraphics[width=4cm]{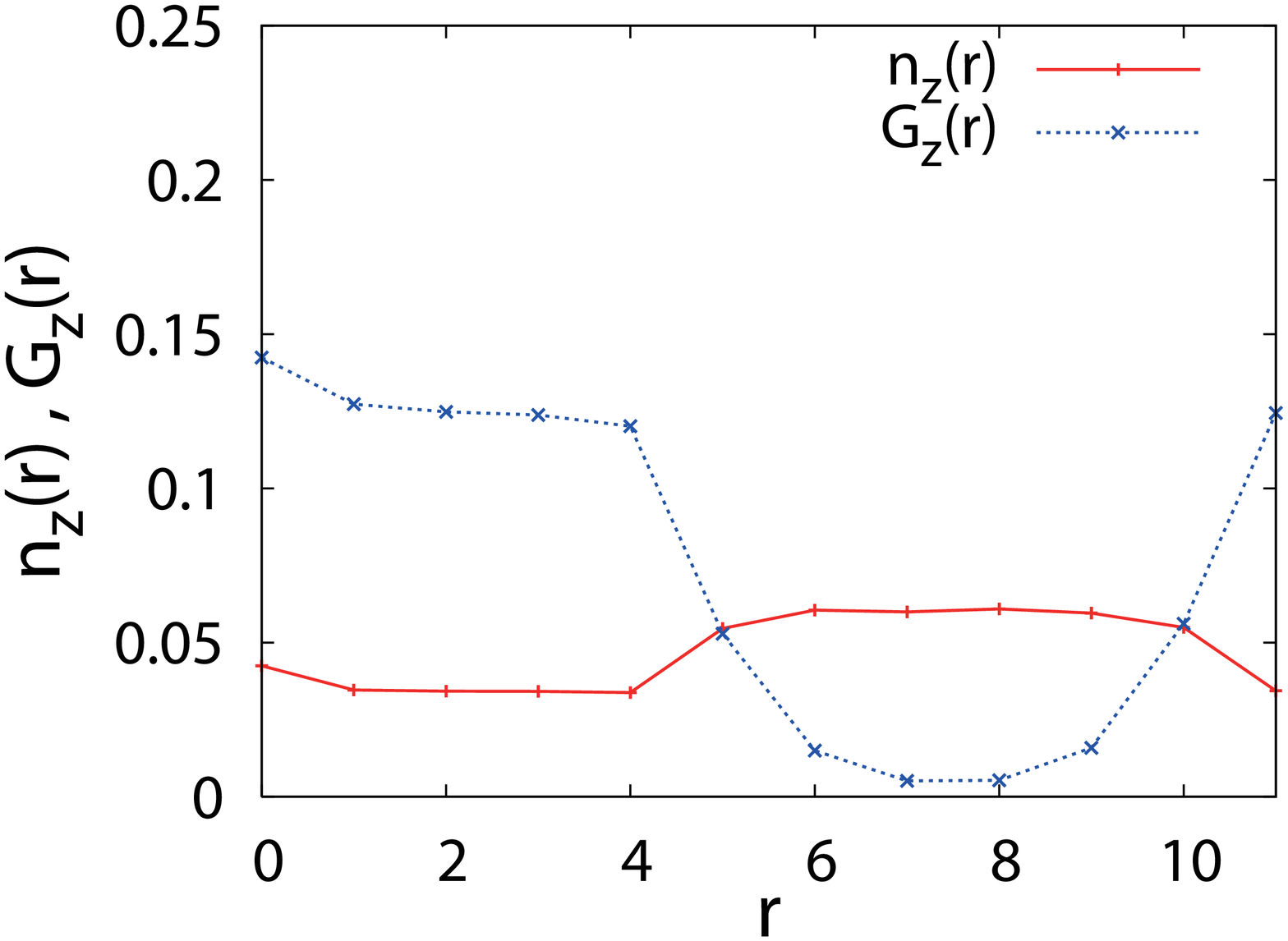} 
\hspace{0.3cm}
\includegraphics[width=4cm]{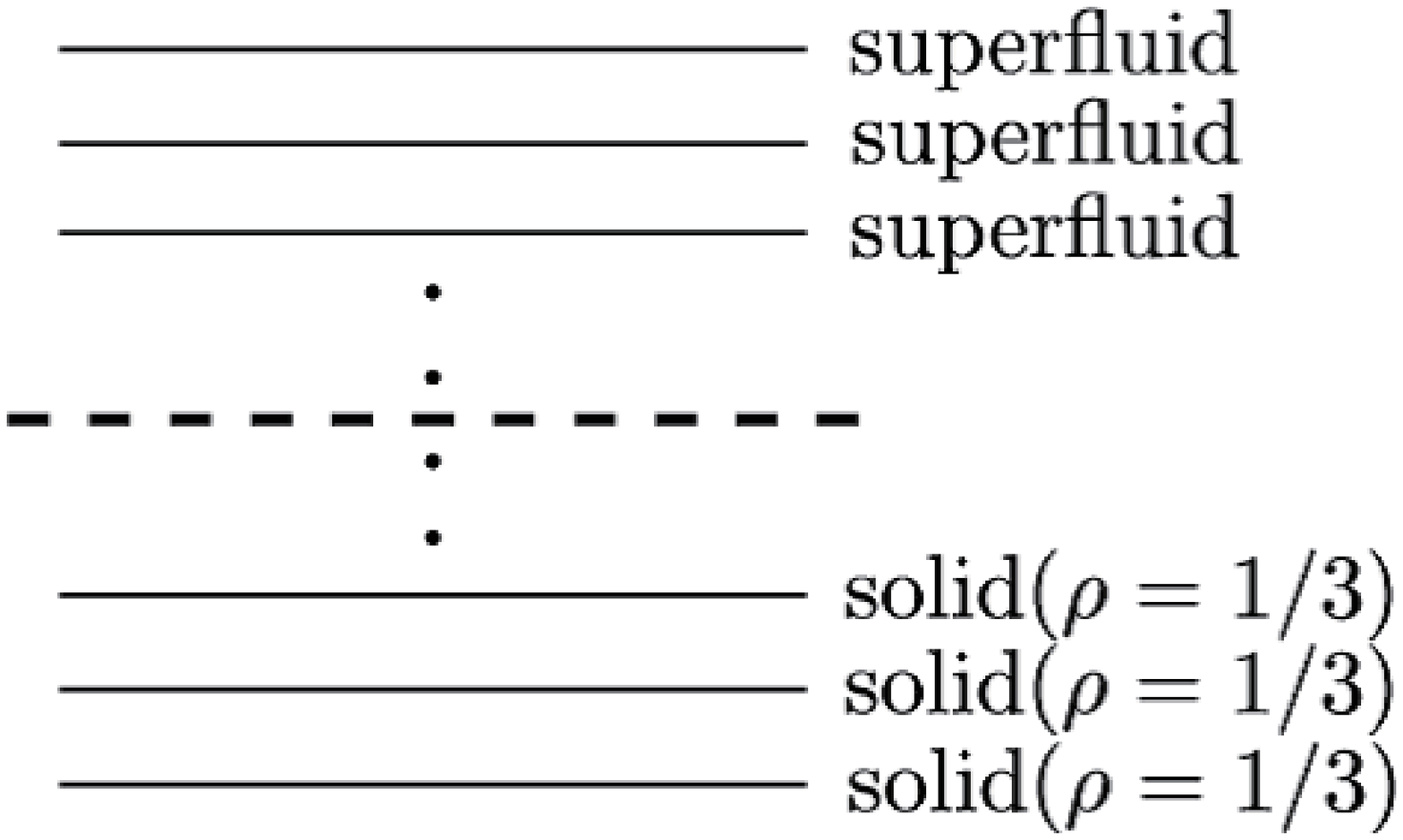}
\caption{
(Left)Number and HC boson correlation functions in the layered direction
in the canonical ensemble for $c_2=50.0$ and $\rho=0.25$.
(Right)Sketch of a PS state in 3D.
}\vspace{-0.7cm}
\label{zcor_canonical}
\end{center}
\end{figure}

It is interesting how the PS state is realized in the present stacked triangular 
lattice\cite{PSsquare}.
In the 2D triangular lattice, the PS state is composed of domains of
the solid and those of the SF.
As shown in Fig.\ref{cor_canonical}, the density and HC boson correlations 
for $c_1=7.0$
seem to imply that the SS is realized there, i.e., the solid-like behavior
of $n(r)$ and the SF-like behavior of $G(r)$.
Then we studied the correlations in the layered direction, and show the 
results in Fig.\ref{zcor_canonical}.
The results indicate that SF layers and solid layers appear separately
as depicted in Fig.\ref{zcor_canonical} in contrast to the 2D case.

In conclusion, we have studied finite-$T$ phase diagram of the HC boson
system in the stacked triangular lattice.
There exist various phases in the phase diagram and the obtained results
are consistent with the phase structure of the HC boson system in the
triangular lattice at $T=0$.
We also found that the SS first loses the superfluidity and then the
density-wave order as $T$ is raised.

\acknowledgments 
This work was partially supported by Grant-in-Aid
for Scientific Research from Japan Society for the 
Promotion of Science under Grant No23540301.


\end{document}